\documentclass[aip, rsi, amsmath, amssymb, reprint]{revtex4-1}
\usepackage{upgreek}
\usepackage{xcolor}
\usepackage{graphicx}
\usepackage{float}
\usepackage{hyperref}
\usepackage{xspace}
\hypersetup{
    colorlinks=true,
    linkcolor=red,
    citecolor=blue,
    filecolor=magenta,
    urlcolor=magenta,
    bookmarks=true
}

\begin{document}


\title{Lithium-Niobate-on-Insulator Waveguide-Integrated Superconducting Nanowire Single-Photon Detectors}

\author{Ayed Al Sayem}
\affiliation{Department of Electrical Engineering, Yale University, New Haven, CT 06511, USA}
\author{Risheng Cheng}
\affiliation{Department of Electrical Engineering, Yale University, New Haven, CT 06511, USA}
\author{Sihao Wang}
\affiliation{Department of Electrical Engineering, Yale University, New Haven, CT 06511, USA}
\author{Hong X. Tang}
\email{hong.tang@yale.edu}
\affiliation{Department of Electrical Engineering, Yale University, New Haven, CT 06511, USA}

\begin{abstract}
We demonstrate waveguide-integrated superconducting nanowire single-photon detectors on thin-film lithium niobate (LN). Using a 250\,$\upmu$m-long NbN superconducting nanowire lithographically defined on top of a 125\,$\upmu$m-long LN nanowaveguide, on-chip detection efficiency of 46\% is realized with simultaneous high performance in dark count rate and timing jitter. As LN possesses high second-order nonlinear $\chi^{(2)}$ and electro-optic properties, an efficient single-photon detector on thin-film LN opens up the possibility to construct small-scale fully-integrated quantum photonic chip which includes single-photon sources, filters, tunable quantum gates and detectors.

\end{abstract}



\maketitle

Waveguide-integrated superconducting nanowire single-photon detectors (SNSPDs) are powerful components that can be exploited to analyze photonic quantum states in an integrated quantum photonic circuit \cite{guo2017parametric,khasminskaya2016fully}. Originally, such detectors have been proposed as an alternative to traditional normal incident planar SNSPDs \cite{gol2001picosecond,marsili2013detecting,cheng2016self,cheng2016large}. Waveguide-integrated SNSPDs rival traditional detectors in terms of efficiency, compactness, dark count rate and timing characteristics \cite{sprengers2011waveguide,pernice2012high, kahl2015waveguide,vetter2016cavity} with added benefit of being compatible with photonic circuit fabrication\cite{guo2017parametric, cheng2019broadband}. In waveguide-integrated SNSPDs, photons are absorbed by thin-film superconducting nanowires situated atop the waveguide through the evanescent coupling \cite{sprengers2011waveguide,kahl2015waveguide,vetter2016cavity,pernice2012high}. Such detectors approach unity efficiency by increasing the nanowire length \cite{pernice2012high}. Although various material platforms such as GaAs\cite{sprengers2011waveguide,sahin2013waveguide}, silicon\cite{munzberg2018superconducting} and SiN \cite{cheng2019broadband,pernice2012high} have been employed to demonstrate such integrated detectors, a much anticipated material platform has been lithium niobate (LiNbO$_3$, LN) which is among  the most desirable photonic materials for classical and quantum integrated optics because of its strong second-order nonlinear $\chi^{(2)}$ and electro-optic properties. Only recently has it been possible to fabricate integrated thin-film LN photonic devices such as modulators \cite{wang2018integrated, wang2018nanophotonic}, high-Q optical resonators \cite{zhang2017monolithic, li2019photon}, low loss non-linear waveguides \cite{lu2019octave}, high-efficiency on-chip periodically poled waveguides \cite{wang2018ultrahigh,jiang2018nonlinear} and rings \cite{chen2019ultra,lu2019periodically}. With the recent experimental demonstration of ultra-high efficiency second harmonic generation \cite{wang2018ultrahigh,chen2019ultra,lu2019periodically}, it is expected that highly efficient spontaneous parametric down-conversion (SPDC) process on an integrated chip with very large signal(idler) to pump photon ratio is now feasible with current fabrication technology. As a result, it opens up the possibility of multi-qubit integration in a single nano-photonic chip \cite{schwartz2018fully,guo2017parametric,paesani2019generation,silverstone2015qubit}. The achievement of such a feat, however, is dependent on efficient single-photon detectors on thin-film LN wavguides which has not been demonstrated so far.

In this Letter, we demonstrate efficient waveguide-integrated SNSPDs on thin-film LN with on-chip detection efficiency (OCDE) of 46\%, dark count rate of 13\,Hz, timing jitter of 32\,ps and noise equivalent power (NEP) of $\mathrm{1.42\times10^{-18}}$\,W/$\mathrm{\sqrt{Hz}}$. Together with the ultralow loss characteristics of thin-film LN waveguides, large second-order non-linear $\chi^{(2)}$ and  electro-optic properties of LN, this work opens the possibility to generate, process and measure quantum light in a compact circuit platform, which altogether impact quantum communication \cite{sibson2017chip} and quantum computation technologies \cite{paesani2019generation,lenzini2018integrated,reimer2019high}. 

\label{sec:examples}
The device is fabricated from a commercial thin-film LN on insulator (LNOI) wafer (supplied by NANOLN) with 615\,nm-thick $z$-cut LN film bonded to  $2\,\upmu\mathrm{m}$-thick silicon dioxide ($\mathrm{SiO_{2}}$) on a $400\,\upmu\mathrm{m}$-thick silicon handle. The photonic waveguiding components including the grating couplers, waveguides, splitters are patterned with electron beam lithography using FOx-16 hydrogen silsesquioxane (HSQ) resist. The patterns are transferred to the LN thin film using an inductively coupled plasma (ICP) reactive ion etching (RIE) tool employing $\mathrm{Ar}^{+}$ plasma.\cite{wang2018ultrahigh} $400\,\mathrm{nm}$ of LN material is etched. A thin oxide layer (10\,nm thickness) is deposited on top of the fabricated photonic devices in order to protect LN from the adverse effect of the subsequent NbN etching by $\mathrm{CF_{4}}$ chemistry. Next, a thin layer of NbN (5\,nm effective thickness) is deposited by the technique of plasma-enhanced atomic layer deposition (PEALD) \cite{cheng2019superconducting}. The transition temperature and sheet resistance of the deposited NbN film are measured to be around 8\,K and 410$\,\Upomega$/sq, respectively. The U-shaped nanowire detectors are patterned using  FOx-16 hydrogen silsesquioxane (HSQ) resist and subsequently etched by $\mathrm{CF_{4}}$ chemistry. Gold electrode pads are fabricated using conventional PMMA and liftoff process.

\begin{figure}[!htbp]
\centering
  \includegraphics[width=8.5cm]{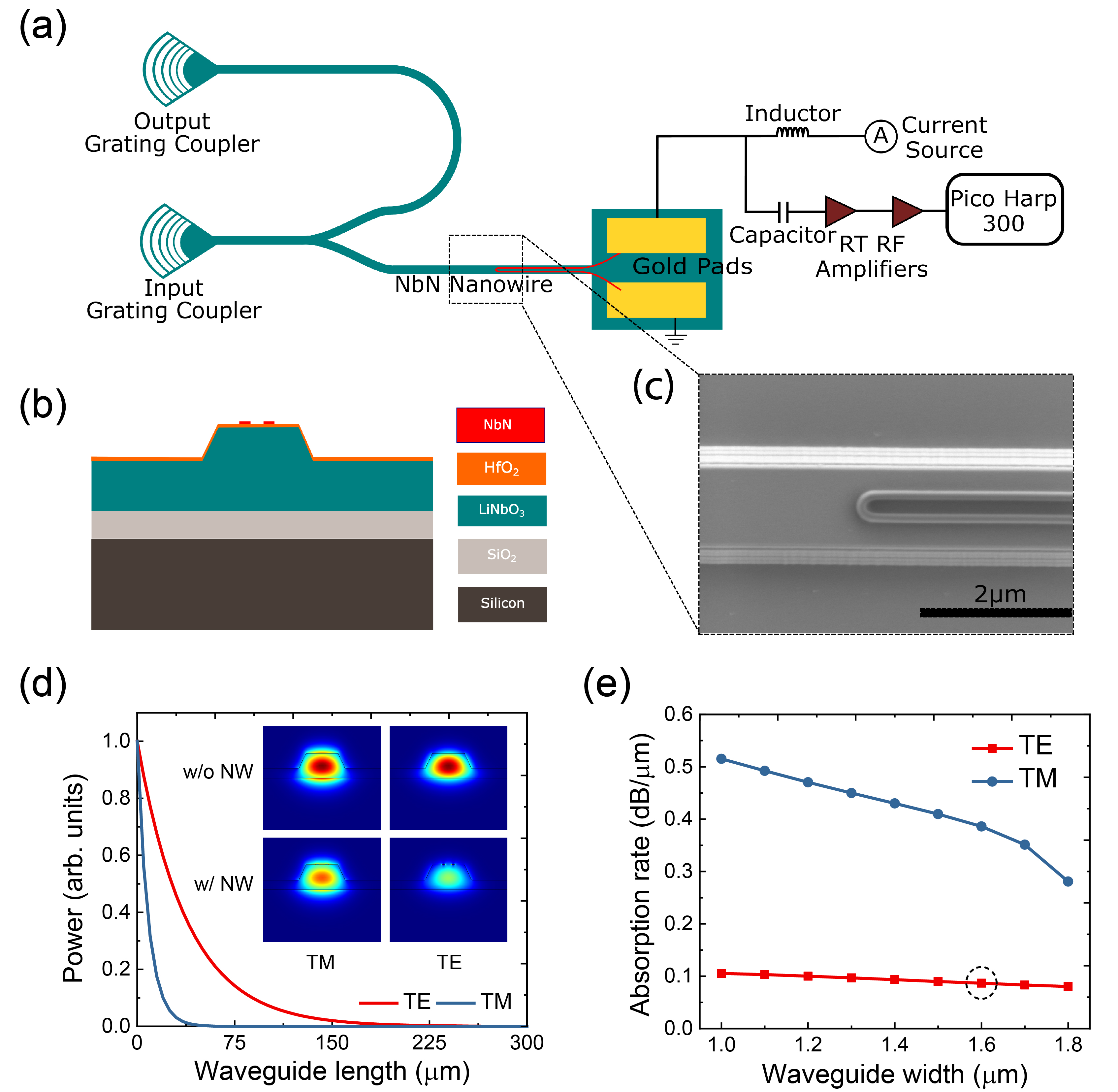}
    \caption{(a) Schematic representation of the device. An input grating coupler is used to couple light into the LN waveguide which couples to U-shaped NbN nanowire. Two Y-splitters and an output grating coupler is used to calibrate the OCDE. (b) Schematic of the cross section. (c)  Scanning electron microscope (SEM) image of the fabricated U-shaped NbN nanowire on top of the LN waveguide. (d) Attenuation as a function of waveguide length (with NbN Nanowire on top) for both TE and TM mode. Inset shows the electric field profiles for both with and without nanowires. (e) Absorption rate as a function of the waveguide width for both TE and TM modes.
    }
\label{Sch}
\end{figure}
Figure~\ref{Sch}(a) shows the schematic of the device. An input grating coupler is used to couple light from the fiber array into the waveguide. The U-shaped NbN nanowire is placed on top of the waveguide. A current source (Keithley 2401) is used to bias the nanowires through gold pads. Two room temperature RF amplifiers (Mini-Circuits ZFL-1000N+) are used to amplify the output pulses from the nanowire detector, and a pulse counter (PicoHarp 300) is employed for the photon counting measurement. To calibrate the OCDE, two Y-splitters and another output grating coupler are used as described later. Figure~\ref{Sch}(b) shows the schematic cross section of the device. Figure~\ref{Sch}(c) shows the scanning electron microscope (SEM) image of the U-shaped NbN nanowire on top of the fabricated LN waveguide. The U-shaped nanowire is positioned almost exactly at the center of the waveguide which ensures high absorption and hence high detection efficiency.

We first analyze the nanowire absorption rate by two dimensional finite-element method (COMSOL) simulation. Figure~\ref{Sch}(d) shows the optical power attenuation as a function of the waveguide length with U-shaped NbN nanowires on top. The mode profiles for TE and TM mode (both with and without nanowire) for a partially etched LN waveguide are shown in the inset. Such LN waveguides (without nanowire) possess ultra-low propagation loss of $\sim$0.2\,dB/cm \cite{lu2019octave,zhang2017monolithic}. The NbN nanowires are 100\,nm wide with 200\,nm spacing between the two nanowires. 
Figure \ref{Sch}(e) shows the absorption rate as a function of the waveguide width for both TE and TM modes. The dashed circle indicates the waveguide width (1.6\,$\upmu$m) used in our experiment. TM mode is less tightly confined than the TE mode. As a result, the evanescent tail of the guided optical mode is coupled more strongly for TM mode than TE mode and has significantly increased absorption rate. From simulation results, it can be calculated that 230\,$\upmu$m (45\,$\upmu$m)-long nanowire is required to achieve 99\% absorption for TE (TM) mode when the waveguide width is 1.6\,$\upmu$m. Though TM modes should provide better results than TE mode, in this experimental demonstration we used TE mode because, for the time being, TE-mode grating couplers can be more robustly designed and patterned. 
The performance of the waveguide-integrated SNSPDs can be characterized by OCDE, dark count rate (DCR), timing jitter, noise equivalent power (NEP) and recovery time. In the following, we characterize each parameter for the fabricated devices. 

\begin{figure}[!htbp]
\centering
 \includegraphics[width=7cm]{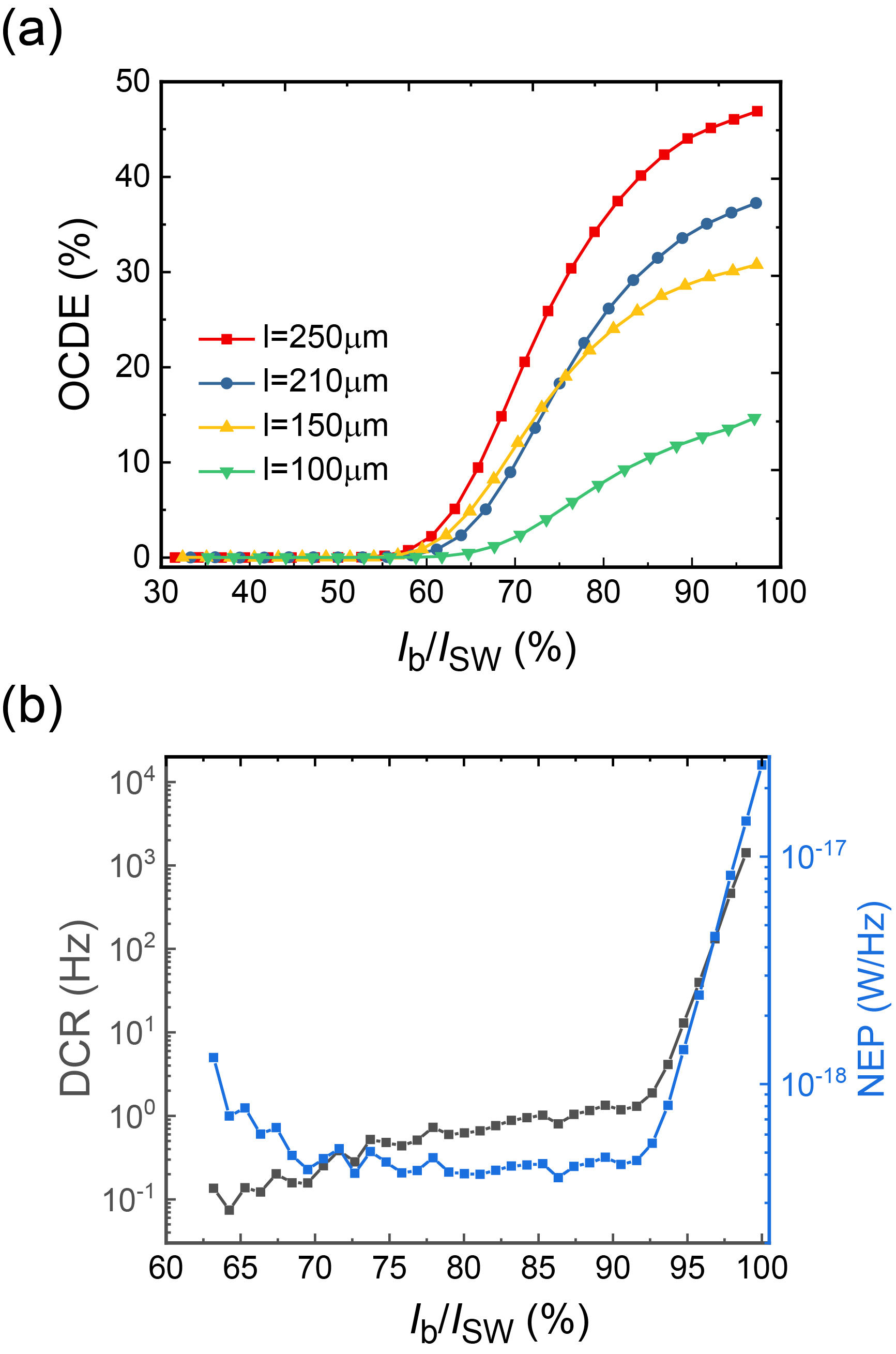}
\caption{(a) OCDE as a function of normalized bias current with varying  lengths of NbN nanowire. (b) The detector DCR and NEP as a function of normalized bias current for 250\,$\upmu$m long detector. At 95\% $I_\mathrm{SW}$, the dark count rate and NEP are ~13\,Hz and $\mathrm{1.42\times10^{-18}}$\,W/$\mathrm{\sqrt{Hz}}$ respectively.}
\label{fig:OCDE}
\end{figure}

For waveguide-integrated SNSPDs, the OCDE is the most relevant parameter, especially for a quantum circuit where single-photon sources and detectors are supposed to be integrated on a single chip\cite{pernice2012high,guo2017parametric,khasminskaya2016fully}. Figure \ref{fig:OCDE}(a) demonstrates the measured OCDE for nanowires of varying length as a function of the normalized bias current $I_\mathrm{b}/I_\mathrm{SW}$, where $I_\mathrm{b}$ and $I_\mathrm{SW}$ represent the bias current and switching current of the nanowire, respectively. The insertion losses of the input grating coupler and the on-chip Y-splitter are calibrated out by the following procedure. The count rate, ${CR}$ can be written in terms of input power $P_\mathrm{in}$, grating coupler efficiency $\eta_\mathrm{gc}$, beam splitter efficiency $\eta_\mathrm{bs}$, attenuation $\eta_\mathrm{attn}$ as, 
\begin{equation}
\label{eq1}
   CR = \frac{\eta_\mathrm{OC}P_\mathrm{in} \eta_\mathrm{bs}\eta_\mathrm{gc}\eta_\mathrm{attn} }{h\nu} 
\end{equation}
where, $\eta_\mathrm{OC}$ is the OCDE, $h\nu$ is the single photon energy. The power of the output grating coupler can be written as, 
\begin{equation}
\label{eq2}
   P_\mathrm{out}=P_\mathrm{in} \eta_\mathrm{bs}^2\eta_\mathrm{gc}^2
\end{equation}
From Eq. \ref{eq1} and \ref{eq2}, the OCDE is given by,
\begin{equation}
\label{eff}
   \eta_\mathrm{OC}=\frac{h \nu CR}{\eta_\mathrm{attn}\sqrt{P_\mathrm{in}P_\mathrm{out}}}
\end{equation}

\begin{figure}[!htbp]
\centering
     \centering\includegraphics[width=7cm]{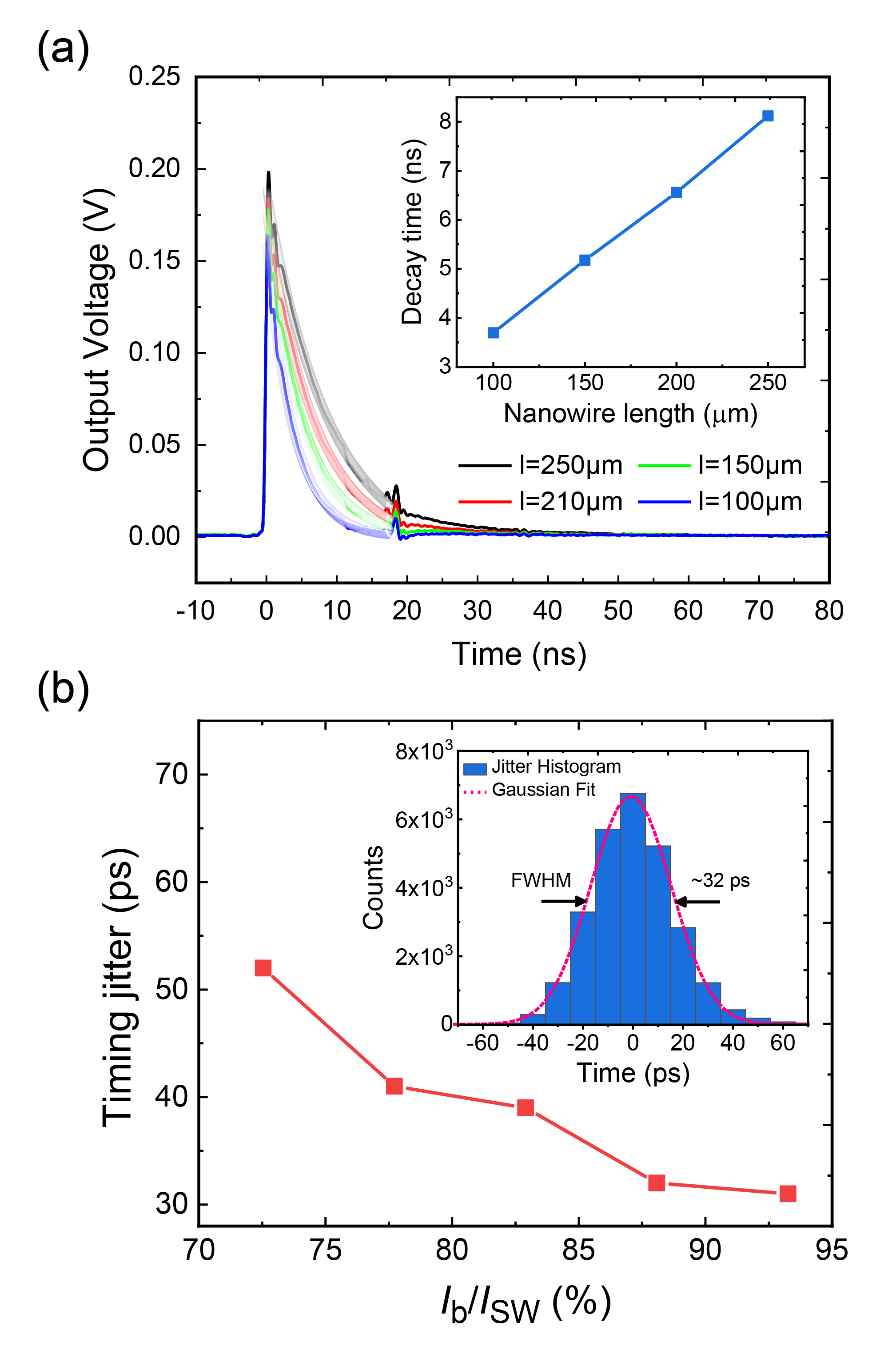}
 \caption{Timing response of the detector (a) Detector output pulse shape with varying  lengths of NbN nanowire. The shaded curves represent the exponential fit. The inset shows the extracted exponential decay time as a function of detector length. (b) Measured timing jitter of the 250\,$\upmu$m-long detector as a function of normalized switching current. The inset shows the histogram of the detector counts at 94\% $I_\mathrm{SW}$. The black curve is the Gaussian fit with an FWHM value of 32\,ps}
      \label{TR}
\end{figure}
From Fig.~\ref{fig:OCDE}(a), it can be observed that longer nanowires provide better efficiency as expected. With 250\,$\upmu$m-long nanowire, the highest OCDE measured is 46\% at 95\% $I_\mathrm{SW}$, which is lower than theoretical calculation. Longer nanowire can in principle improve efficiency but using longer nanowire, the detectors do not saturate as well as shorter nanowires and also the timing jitter gets degraded. We attribute the lowered efficiency mostly due to non-perfect quality of the NbN film deposited on patterned LiNbO$_3$ substrates. It should be pointed out that in contrast to previous work on Si and Si$_3$N$_4$ waveguide-integrated SNSPDs\cite{pernice2012high,schuck2013waveguide,schuck2013nbtin}, where the superconducting thin film is deposited prior to the waveguide fabrication, in the current processing, we resort to the deposition of superconducting films after the formation of the LN waveguide. 

We select the 250\,$\upmu$m-long nanowire detector with the highest OCDE to characterize its DCR as a function of $I_\mathrm{b}/I_\mathrm{SW}$, as shown in Fig.~\ref{fig:OCDE}(b). When biased at 95\% $I_\mathrm{SW}$, the DCR is measured to be around 13\,Hz with ~46\% OCDE. The NEP is dependent on both OCDE and dark count rate ($R_\mathrm{DC}$), $\mathrm{NEP} = h\nu(2R_\mathrm{DC})^{1/2} /\eta_\mathrm{OC}$. Figure \ref{fig:OCDE}(b) shows both the DCR and NEP as a function of bias current. Both OCDE and DCR has been measured at the same condition when when the input fiber is well aligned with the grating coupler. At the bias current of ~90\%\,$I_\mathrm{SW}$ and ~95\%\,$I_\mathrm{SW}$, the NEP is measured to be $\mathrm{1.42\times10^{-18}}$\,W/$\mathrm{\sqrt{Hz}}$ and $\mathrm{4.43\times10^{-19}}$\,W/$\mathrm{\sqrt{Hz}}$, which is fairly close to other waveguide-integrated SNSPDs \cite{pernice2012high,schuck2013nbtin,schuck2013waveguide}. 
For the applications in time-domian multiplexing and optical buffering, SNSPDs with low timing jitter are desired. To characterize timing jitter, we use a femtosecond-pulsed 1560\,nm laser (Toptica FemtoFErb 1560) to illuminate the detector, and the arrival time differences ($\Delta t $) between the photon-excited detector signal and the synchronization trigger signal from the laser are recorded by a 40\,Sample/s real-time oscilloscope (Lecroy HDO9404). Figure \ref{TR}(a) shows the output voltage pulses from four detectors of different nanowire length. The inset shows the extracted decay time from the exponential fitting. Figure \ref{TR}(b) shows the measured jitter of the 250\,$\upmu$m-long detector as a function of $I_\mathrm{b}/I_\mathrm{SW}$ with histogram results shown in the inset. The Gaussian fitting yields the best timing jitter value of 32\,ps defined as full-width at half maximum (FWHM) of the histogram peak. 

Currently, the maximum OCDE is lower than the theoretical prediction. The OCDE can be improved by using narrower (for better saturation) and longer nanowires (for better absorption). One simple way to improve the efficiency is to utilize TM mode instead of TE modes which provides much higher absorption rate. With such modifications, it should be possible to reach more than 90\% OCDE.


In conclusion, we have experimentally demonstrated waveguide-integrated SNSPDs on thin-film LN. Utilizing the high second-order nonlinear $\chi^{(2)}$ and electro-optic properties, we expect such single-photon detectors on thin-film LN waveguides can play a significant role in various integrated photonic platforms such as artificial neural networks, integrated quantum photonic circuits, deterministic heralded single-photon sources by time multiplexing, tunable single-photon spectrometers and so on. 

\section*{Funding Information}
This work is supported by Department of Energy, Office of Basic Energy Sciences, Division of Materials Sciences and Engineering under Grant DE-SC0019406. 

\section*{Acknowledgments}
The facilities used for device fabrication were supported by the Yale SEAS Cleanroom and the Yale Institute for Nanoscience and Quantum Engineering (YINQE). The authors would like to thank Dr. Yong Sun, Dr. Michael Rooks, Sean Rinehart, and Kelly Woods for their assistance provided in the device fabrication.

\bibliography{MAIN}

\end{document}